\documentclass{article}
\usepackage{spconf,amsmath,graphicx}
\usepackage{multirow}
\usepackage{adjustbox}
\usepackage{dblfloatfix}
\usepackage{subcaption}
\usepackage{spconf,amsmath,graphicx}
\usepackage{multirow}
\usepackage{adjustbox}
\usepackage{dblfloatfix}
\usepackage{subcaption}
\usepackage{url,times,booktabs}
\usepackage{comment}
\usepackage{tabularx}
\usepackage{makecell}
\usepackage{cite}
\usepackage{enumerate}
\usepackage{amsmath,graphicx,url,times,booktabs, tabularx}
\usepackage{arydshln}

\title{A STUDY ON THE ROLE OF SUBSIDIARY INFORMATION \\IN REPLAY ATTACK SPOOFING DETECTION}

\name{Jee-weon Jung, Hye-jin Shim, Hee-Soo Heo, and Ha-Jin Yu$^\dag$\thanks{$^\dag$ Corresponding author}\thanks{This research was supported by Basic Science Research Program through the National Research Foundation of Korea(NRF) funded by the Ministry of  Science, ICT \& Future Planning(2017R1A2B4011609)}}
\address{School of Computer Science, University of Seoul, South Korea}

\begin{document}
\maketitle

\begin{abstract}
In this study, we analyze the role of various categories of subsidiary information in conducting replay attack spoofing detection: `Room Size', `Reverberation', `Speaker-to-ASV distance, `Attacker-to-Speaker distance', and `Replay Device Quality'. 
As a means of analyzing subsidiary information, we use two frameworks to either subtract or include a category of subsidiary information to the code extracted from a deep neural network. 
For subtraction, we utilize an adversarial process framework which makes the code orthogonal to the basis vectors of the subsidiary information. 
For addition, we utilize the multi-task learning framework to include subsidiary information to the code. 
All experiments are conducted using the ASVspoof 2019 physical access scenario with the provided meta data.  
Through the analysis of the result of the two approaches, we conclude that various categories of subsidiary information does not reside enough in the code when the deep neural network is trained for binary classification. 
Explicitly including various categories of subsidiary information through the multi-task learning framework can help improve performance in closed set condition. 
\end{abstract}


\section{Introduction}
\label{sec:1. intro}
In recent years, advances in automatic speaker recognition have made audio spoofing attacks possible and created a need for countermeasures to prevent such attacks \cite{snyder2018x, tom2018end, spoofCLDNN,font2017experimental, suthokumar2017independent, xiao2015spoofing, srinivas2018relative, li2018multiple, gunendradasan2018detection}. 
The ASVspoof challenges are providing a common platform for researchers to verify and compare various approaches \cite{wu2015asvspoof, kinnunen2017asvspoof, todisco2019asvspoof}. 
Among such challenges, the 2019 challenge provides two scenarios for the countermeasures against audio spoofing attacks with controlled configurations for research: the logical and physical access (PA) scenarios. 
In this study, we concentrate on the physical scenario, which is also referred to as replay attack detection. 
The task is a binary classification problem where an input audio is classified into either \textit{bona-fide} (also called genuine) or \textit{spoofed} (replayed). 

Recently, deep neural networks (DNNs) have displayed promising performances across a  range of academic and industrial tasks, including the image and audio domains \cite{zagoruyko2016wide,Residual}. 
One of the research objective of DNN-related studies is to train DNN as a feature extractor, where a N-dimensional representation vector from the last hidden layer is used as the feature for the target task (i.e. d-vector/x-vector speaker embeddings for speaker recognition, referred to as `\textit{code}' throughout this paper) \cite{d-vector, snyder2018x, shon2018frame, DeepSpeaker}. 
Regarding this research objective, a number of studies are devising techniques to either subtract certain information from or incorporate it in the representation vector \cite{heo2018cosine, MultitaskLearning}. 
For example, Heo \textit{et al.} \cite{heo2018cosine} devised an adversarial scheme called cosine adversarial network (CAN) to train the code to become orthogonal to the basis vectors of the subsidiary information which is known to be an obstacle for the target task. 
As an another example, Chen \textit{et al.} \cite{Multi-taskLearning2} used a multi-task learning (MTL) framework \cite{MultitaskLearning} to incorporate phoneme information to the code for the speaker verification task with the hypothesis that the code which also includes phoneme information would better represent the input utterance. 

In this study, we further analyze the role of various categories of subsidiary information in replay attack spoofing detection. 
Specifically, we determine whether each category of subsidiary information is beneficial for spoofing detection by observing the loss and equal error rates (EERs) when each category of subsidiary information is either removed or added. 
For this goal, we use two frameworks introduced in the previous paragraph: CAN for subtracting and MTL for adding various categories of subsidiary information. 
We analyze all five categories of subsidiary information (`Room Size', `Reverberation', `Speaker-to-ASV distance', `Attacker-to-Speaker distance', and `Replay Device Quality') provided in the ASVspoof 2019 PA scenario dataset. 

Our study differs from the conventional studies that deal with subsidiary information in the following two ways \cite{shim2018replay,lavrentyeva2017audio,nagarsheth2017replay}.
First, we analyzed the individual effects of each category of subsidiary information separately, while previous studies merely included combinations of different categories of subsidiary information to the code. 
Second, we analyzed the role of each subsidiary information in more detail, whether such subsidiary information is required for conducting replay spoofing detection. 
We did this by including and by trying to exclude it from the code. 

Through the analysis regarding the role of various categories of subsidiary information we show that all five categories of subsidiary information does not reside in the code of the conventional systems enough through experiments that subsidiary information classification cannot be conducted using the code, trained for replay attack detection in a in binary classification task (bona-fide/spoofed). 
By including various categories of subsidiary information to the code, we found performance improvements in replay spoofing detection in closed set condition. 

The rest of this paper is organized as follows: Section 2 introduces the baseline end-to-end (E2E) DNN model used for replay attack spoofing detection. 
The two frameworks, for subtracting and adding a category of subsidiary information, are addressed in Section 3. 
The experiments and result analysis are presented in Section 4 and 5 respectively. 
The paper is concluded in Section 6.

\begin{table}[t]
 \caption{Architecture of the E2E DNN. Each number in `Output shape' refers to frame (time), frequency, and the number of filters respectively. The number 120 of `Input' is only for training phase for mini-batch construction. At test phase, utterance with varying duration is input to the model. All convolutional layers have filter length of 3 for frame and 7 for frequency dimension. Three numbers inside the bracket of convolutional layers refer to stride size in frame and frequency dimension and the number of filters. Batch normalization and the first activation is omitted at the first residual block following \cite{he2016identity}.}
  \centering
  \label{tab:table1}
  \begin{tabular}{l c c}
  \toprule
  \textbf{Layer} & \textbf{Input:(120, 1025)} & \textbf{Output shape}\\
  \Xhline{2\arrayrulewidth}
  \multirow{3}{*}{Conv1} & Conv(3,3,128) & \multirow{3}{*}{(120, 1025, 16)}\\
  \multirow{2}{*}{}& BN & \\
  & LReLU & \\
  \midrule
  Res block & 
    $\left \{
      \begin{tabular}{c}
      BN \\
      LReLU\\
      Conv(2, 4, 128)\\
      BN\\
      LReLU\\
      Conv(2, 4, 128)\\
      \end{tabular}
    \right \}$
    $\times$9
    
  & (15, 17, 128)\\
  \midrule
  AvgPool & Pool(1, 17) & (15, 128)\\
  \midrule
  GRU & GRU(512) & (512,)\\
  \midrule
  Code & FC(64) & (64,)\\
  \midrule
  Output & FC(2) & (2,)\\
  \bottomrule
  \end{tabular}
\end{table}

\begin{figure}[t]
\begin{center}
\begin{subfigure}{\linewidth}
\centering
\includegraphics[width=\linewidth]{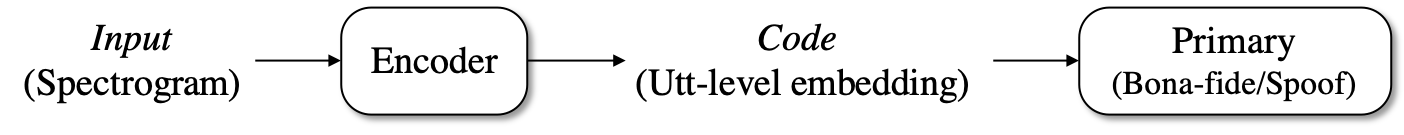} 
\subcaption{}
\label{fig:1_1}
\end{subfigure}

\begin{subfigure}{\linewidth}
\includegraphics[width=\linewidth]{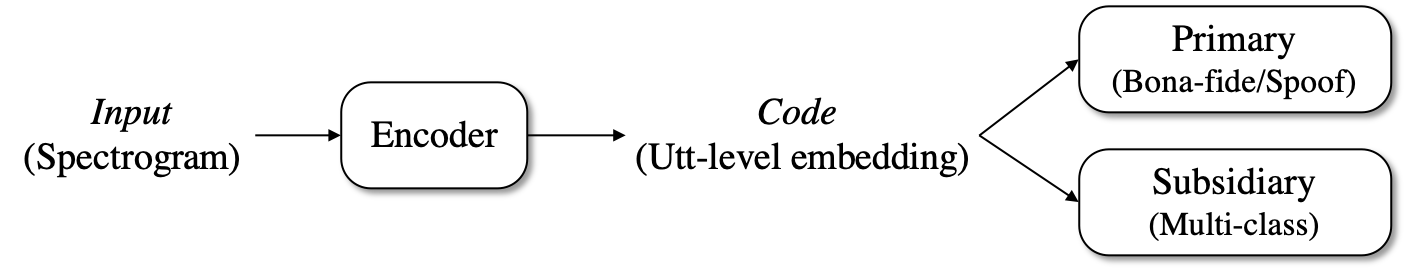} 
\subcaption{}
\label{fig:1_2}
\end{subfigure}
 
\caption{Simplified architecture of the E2E DNN, CAN, and MTL. In the CAN framework, code is trained to become orthogonal to basis vectors of the subsidiary model by the adversarial process. In the MTL framework, code is trained to include both information regarding the primary model and the subsidiary model. (a) E2E DNN described in Table \ref{tab:table1} is interpreted as a combination of encoder and primary model. (b) Architecture of CAN and MTL. For CAN, subsidiary model comprises 2 more hidden layers and an output layer following that of \cite{heo2018cosine}. For MTL, subsidiary model comprises only the output layer.}
\label{fig:1}
\end{center}
\end{figure}

\section{End-to-end DNN Baseline}
\label{sec:2. e2e base}
\subsection{Model architecture}
End-to-end (E2E) DNN is a promising architecture for various tasks \cite{generalizedE2E, jung2018complete, tom2018end}. 
It incorporates both feature extraction and classification into a single DNN to fully exploit the merit of a data-driven training scheme. 
In this study, we use the E2E DNN of \cite{jung2019Replay} which showed promising performance at the ASVspoof 2019 PA challenge despite its simple process pipeline \footnote{A Github link with the full code of our implementation of the model will be released after the anonymity period ends}. 

Table \ref{tab:table1} describes the E2E DNN architecture used in this study.
This network takes spectrograms as input, and outputs the result of replay attack detection using an output layer with two nodes indicating bona-fide and spoofed, respectively. 
We used residual blocks that comprise convolutional layers, batch normalization layers \cite{BatchNormalization}, and leaky rectified linear unit (LReLU) activation functions \cite{leaky} following the identity mapping \cite{he2016identity} of He \textit{et al.} used to extract multiple segment-level embeddings from an input spectrogram. 
Then, the frequency axis is averaged using an average pooling layer. 
A gated recurrent unit (GRU) layer then extracts an utterance-level embedding followed by a fully-connected layer (the code). 

\begin{figure*}[ht!]
\begin{center}

\begin{subfigure}{0.33\textwidth}
\centering
\includegraphics[width=\textwidth]{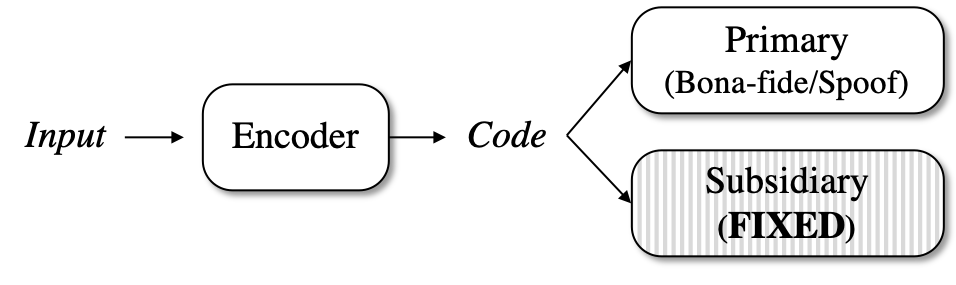} 
\subcaption{}
\label{fig:2_1}
\end{subfigure}
\begin{subfigure}{0.33\textwidth}
\centering
\includegraphics[width=\textwidth]{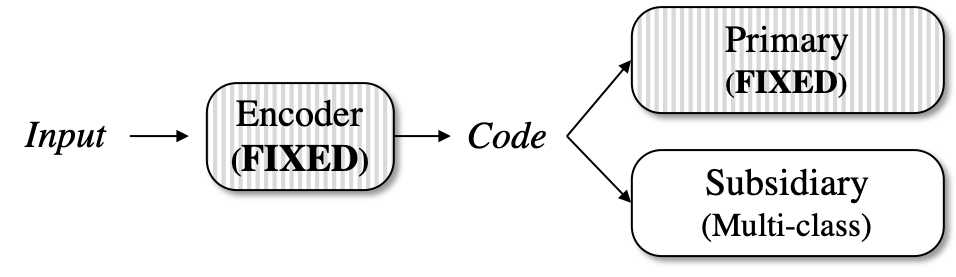} 
\subcaption{}
\label{fig:2_2}
\end{subfigure}
\begin{subfigure}{0.33\textwidth}
\centering
\includegraphics[width=\linewidth]{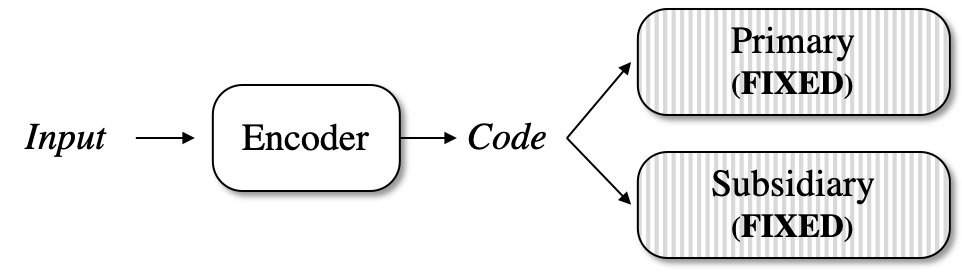} 
\subcaption{}
\label{fig:2_3}
\end{subfigure}
 
\caption{Three phases for the training of CAN framework. These phases are repeated, starting from (a). (a) The encoder and the primary model are trained, equivalent to E2E DNN model training. (b) The encoder and the primary model are frozen, and the subsidiary model is trained. The amount of subsidiary information that resides in the current encoder is evaluated. (c) Both primary and subsidiary models are frozen, and the encoder is trained. Here, the encoder is trained to produce the code that is orthogonal to the basis vectors of the subsidiary model.}
\label{fig:2}
\end{center}
\end{figure*}

\subsection{Ring loss}
A number of loss functions are being studied to supplement the softmax output \cite{wen2016discriminative,zheng2018ring}. 
Among these loss functions, Ring loss \cite{zheng2018ring}, which was first proposed for the face recognition task, has been showing promising results. 
Ring loss normalizes the code (from the last hidden layer) to a learnt value $\mathcal{R}$ by penalizing how far the norm of the code is from the current $\mathcal{R}$. 
Ring loss can be defined by:
\begin{equation}
    \mathcal{L}_\mathcal{R} = \frac{\lambda}{2m}\sum(||\mathcal{F}(x_i)||_2-\mathcal{R})^2,
\end{equation}
where $\lambda$ refers to the weight of the Ring loss (weight of CCE loss is assumed to be 1), $m$ refers to the mini-batch size, $\mathcal{R}$ refers to the radius of the ring (norm value) and $\mathcal{F}(x_i)$ is the code. 
In this study, we compare experimental results of both with and without Ring loss (see Section 5).

\section{Frameworks for exploiting\\subsidiary information}
\label{sec:3. Frames}
In this section, we introduce two frameworks for subtracting and adding a certain category of subsidiary information: cosine adversarial network (CAN) \cite{heo2018cosine} and multi-task learning (MTL) \cite{MultitaskLearning}. 
To introduce both frameworks, we reinterpret the E2E DNN as illustrated in Figure \ref{fig:1}. 
In this point of view, the E2E DNN explained in the previous section is interpreted as a combination of an \textit{`encoder'} which extracts the code from spectrogram and a \textit{`primary'} model which conducts binary classification using the code. 

\subsection{Cosine Adversarial Network}
Cosine adversarial network (CAN) was proposed to eliminate subsidiary information from the code extracted by an encoder \cite{heo2018cosine}. 
By eliminating subsidiary information that is known to decrease the performance of the target task, improvements in performance have been observed.  
For example, improved performance was shown by removing the channel information in the speaker verification. 

In this study, we use the CAN framework to examine whether various categories of subsidiary information are included in the code for conducting replay spoofing attack detection when the model is trained for a binary classification task. 
Five categories of subsidiary information are analyzed using the meta data provided from the ASVspoof 2019 PA dataset: `Room Size', `Reverberation', `Speaker-to-Mic distance', `Attacker-to-Speaker distance', and `Replay Device Quality'.

The CAN framework is trained using a repetition of the three phases shown in Figure \ref{fig:2}. 
In the first phase, the parameters of the subsidiary model are frozen, and the encoder and the primary model are trained. 
This procedure is equivalent to the training of E2E DNN introduced in Section 2. 
In the second phase, parameters of the encoder and the primary model are frozen, and the subsidiary model is trained. 
If the loss decreases well in this process, it can be interpreted that the subsidiary information actually exists in the current code, and if the loss does not decreases well, it can be the first case which explained in the previous paragraph (subsidiary information does not reside in the code). 
In the third phase, parameters of the primary model and the subsidiary model are frozen, and the encoder is trained to exclude subsidiary information by the adversarial process. 
Here, we remove the subsidiary information by the encoder by training the code to be orthogonal from basis vectors of the subsidiary model, trained at the previous phase. 
By repeating these three phases, CAN trains the DNN to perform the primary task and exclude subsidiary information.

Here, interpret the results of the CAN in three cases as follows.
\begin{enumerate}
    \item The encoder and the primary model is successfully trained, but the subsidiary loss does not decrease while the parameters of encoder is frozen: subsidiary information does not reside in the code enough to conduct subsidiary information classification when the DNN is trained as a binary classifier. 
    \item The CAN framework is successfully trained, but the performance of the replay detection decrease as subsidiary information is excluded: subsidiary information is helpful for conducting replay detection.
    \item The CAN framework is successfully trained, and performance is increased: subsidiary information is an obstacle for replay attack detection. 
\end{enumerate}

\subsection{Multi-task Learning Framework}
The multi-task learning (MTL) framework was proposed by Caruana \textit{et al.} to train a DNN that can conduct more than one task \cite{MultitaskLearning} simultaneously. 
In a number of studies, this framework has successfully incorporated additional information into the code by adding additional output layers. 
By training with more than a single task, better generalization performance has also been reported which is interpreted as a result of the network not being overfitted to a single task. 
For example, Chen \textit{et al.} reported better performance in speaker verification by explicitly training phoneme information with the speaker information \cite{Multi-taskLearning2}. 
In replay attack detection, \cite{shim2018replay} showed that the information regarding the subsidiary task could be included in the code and improves the performance on the ASVspoof 2017 dataset. 
However, \cite{shim2018replay} used a combination of various categories of subsidiary information that does not reveal the role of each category of subsidiary information. 
In this study, we analyze the effect of adding each category of subsidiary information via the MTL framework. 

The MTL framework is trained using the following equation:
\begin{equation}
    \mathcal{L} = \lambda_{pri} \mathcal{L}_{pri} + \lambda_{sub} \mathcal{L}_{sub},
\end{equation}
where $\mathcal{L}$ refers to the final loss, $\mathcal{L}_{pri}$, $\mathcal{L}_{sub}$ refer to the categorical cross-entropy (CCE) loss, and $\lambda_{pri}$, $\lambda_{sub}$ refer to their weights respectively. 
Comparing to the CAN framework, the MTL framework can be interpreted as training the encoder, the primary, and the subsidiary model concurrently.

\subsection{Modified MTL Framework for Replay Spoofing Detection}
When the MTL framework is utilized in replay attack spoofing detection for adding subsidiary information, overlaps can occur between different tasks. 
Using `Replay Quality' information, for example, when bona-fide utterance is input, there are no labels for the subsidiary task. 
In the authors' previous study, these confusions between the primary task and the subsidiary task were further analyzed and a novel scheme was proposed\footnote{Referred to as authors' work for now, but will cite papers after the anonymity period ends}. 

In the modified scheme, there exists only one output layer. 
The number of nodes of the output layer is equal to the number of defined subsidiary task's labels  $+ 1$, where the additional node is for bona-fide input. 
Through comparison experiments conducted in the authors' previous study using the ASVspoof 2017 dataset \cite{kinnunen2017asvspoof}, this scheme clearly outperformed that of the original MTL framework. 
Comparison experiments of this scheme with the MTL framework on the ASV2019 PA dataset are described in Table \ref{table:modifiedMTL}.
Note that because the used architecture is E2E, we use the value of the bona-fide node directly as the score.

\section{Experimental Settings}
\label{sec:4. exp}
All experiments of this paper were conducted using PyTorch \cite{paszke2017automatic}, a deep learning library in Python\footnote{A Github link to the full experimental code will be released after the anonymity period ends}. 

\subsection{Dataset}
We used the ASVspoof 2019 PA dataset for all experiments \cite{todisco2019asvspoof}. 
Table \ref{table:dataset} describes the subset configuration. 
This dataset comprises 20 speakers (8 male, 12 female) and all utterances are recorded at 16-kHz sampling rate with 16-bit resolution. 
The training subset comprises 27 different acoustic configurations in total which both bona-fide and replayed utterances share: a combination of three `Room Size', three `Reverberation', and three `Speaker-to-ASV Mic distance'. 
Nine different replay configurations are used which only replayed utterances have: a combination of three `Attacker-to-talker distance' and three `Replay Device Quality'. 
Note that the development set is a closed set and the evaluation set is an open set configuration (e.g. utterances in the development set will share the level of `Reverberation' with the train set, and the unknown level of `Reverberation' will be included in the evaluation set). 

\begin{table}[t]
  \caption{Subset configuration of the ASVspoof 2019 physical access dataset.}
  \label{tab:table2}
  \centering
  \begin{tabular}{lcc}
  \toprule
  \multirow{2}{*}{\textbf{Subset}} & \multicolumn{2}{c}{\textbf{\# utterances}}\\
  & bona-fide & spoof\\
  \midrule
  Training & 5,400 & 48,600\\
  Development & 5,400 & 24,300\\
  Evaluation & 18,090 & 119,367\\
  \bottomrule
  \end{tabular}
\label{table:dataset}
\end{table}

\subsection{Acoustic Feature}
Magnitude spectrograms with 2048 points fast Fourier transform were used for all experiments in this study. 
The window length and the shift size were 50 ms and 30 ms respectively, following \cite{jung2019Replay}. 
Normalization on mean or standard deviation was not applied. 
In the training phase, 120 randomly selected continuous frames were used for mini-batch construction. 
In the test (inference) phase, entire frames were used. 

\subsection{Experimental Settings: Common for Baseline E2E DNN, CAN and MTL}
The encoder consists of nine residual blocks, a GRU layer, and a code layer. 
The encoder is common to E2E DNN, CAN and MTL. 
Each residual blocks have filter size of (3, 7) and stride size of (2, 4) where inequality in the size was set considering the size of input spectrogram ((120, 1025) for training). 
The GRU layer has 512 nodes. 
The code layer is fully-connected layer with 64 nodes. 
We used LReLU \cite{leaky} for all the non-linearity.

We used the AMSGrad \cite{reddi2018convergence} optimizer with a learning rate of 0.0005. 
A weight decay with 0.0001 was applied to all parameters. 
The size of the mini-batch is 32. 
For models with Ring loss, the weights for the Ring loss and CCE loss are the same. 

For each epoch, we made the training set with all 5400 bona-fide utterances and randomly selected 5400 spoofing utterances to balance the number of samples per each class. 
In our internal comparison experiments, this method brought performance improvement. 

The primary model consists of only one output layer (fully-connected, 2 nodes indicating bona-fide/spoof) with no other hidden layers, for all E2E DNN, CAN, and MTL.

\subsection{Experimental Settings: CAN}
For experiments with the CAN framework, the subsidiary model consists of two fully-connected hidden layers with 128 nodes, and an output layer with three nodes following \cite{heo2018cosine}.  
The relative proportions of the first, second, and the third phase, was set to 3:1:1, where the first refers to Figure 2-(a), which train the encoder and the primary model (This is identical to Baseline E2E training). 

\subsection{Experimental Settings: MTL}
For experiments with the MTL framework, the subsidiary model consists of an output layer that has three nodes following \cite{shim2018replay}. 
The weight of the subsidiary loss ($\mathcal{L}_{sub}$ of Eq. 2) was set to 0.5. 
When using replay-related subsidiary information (`Attacker-to-Talker' and `Replay Quality'), we added one node to the subsidiary model's output layer for bona-fide input. 

\section{Results and analysis}

\begin{table*}[t] 
  \caption{Results using the MTL framework on five categories of subsidiary information. Bold indicates the result better than the baseline. Performance reported in EER (\%).}
  \centering
  \makebox[\textwidth][c]{
  \begin{tabular}{l cc cccccccccc}
  \toprule
   & \multicolumn{2}{c}{Baseline E2E} & \multicolumn{2}{c}{Room size} & \multicolumn{2}{c}{Reverberation} & \multicolumn{2}{c}{Spk-to-ASV mic} & \multicolumn{2}{c}{Attacker-to-Talker} & \multicolumn{2}{c}{Replay Quality}\\
   \cmidrule(lr){2-3}
   \cmidrule(lr){4-5}
   \cmidrule(lr){6-7}
   \cmidrule(lr){8-9}
   \cmidrule(lr){10-11}
   \cmidrule(lr){12-13}
   & Val & Eval & Val & Eval & Val & Eval & Val & Eval & Val & Eval & Val & Eval\\
  \midrule
  w/o ring & 1.26&4.79 & 1.41&6.66 & 1.30&\textbf{4.65} & \textbf{1.14}&5.00 & \textbf{1.08}&5.40 & 1.46&\textbf{4.63}\\
  w ring & 1.31&4.53 & \textbf{1.09}&5.25 & \textbf{0.91}&\textbf{4.51} & \textbf{1.04}&4.65 & \textbf{1.31}&5.27 & \textbf{1.22}&4.85\\
  \bottomrule
  \end{tabular}}
\label{table:MTL}
\end{table*}

\begin{table}[t] 
  \caption{Additional experimental results of the modified MTL introduced in Section 3.3. This scheme can be used only for replay-related subsidiary information. Performance reported in EER (\%). For comparison with the `Baseline', refer to Table \ref{table:MTL}}
  \centering
  \begin{tabular}{l cccc}
  \toprule
   & \multicolumn{2}{c}{Attacker-to-Talker} & \multicolumn{2}{c}{Replay Quality}\\
   \cmidrule(lr){2-3}\cmidrule(lr){4-5}
   & Val & Eval & Val & Eval\\
  \midrule
  w/o ring & \textbf{1.13}&6.20 & 1.99&4.80\\
  w ring & \textbf{0.91}&7.03 & 2.11&5.46\\
  \bottomrule
  \end{tabular}
\label{table:modifiedMTL}
\end{table}

\subsection{Removing subsidiary information (CAN)}
Experiments conducted on all five categories showed identical results that the loss of subsidiary task did not decrease while the encoder was frozen (The first case among the three described in Section 3.1.). 
We interpreted this result as that the subsidiary information does not \textit{naturally} reside enough in the code to conduct subsidiary information classification when the DNN is trained using a binary classification scheme (bona-fide/replayed). 
In other words, training a spoofing detector using a simple binary classification scheme could not utilize the subsidiary information. 
Based on this interpretation, the results in \cite{shim2018replay, lavrentyeva2017audio} which have seen improvement in performance by multi-task learning means that adding subsidiary information can enhance the discrimination power of the code. 

We omit the table for the performance of CAN framework because we concluded that subsidiary information does not reside in the code enough to conduct the classification of subsidiary information, which the CAN framework aims to remove, making the EER meaningless.
Note that for all five categories of subsidiary information, removing subsidiary information with the CAN framework worsened the performances in terms of EER of the primary task (replay attack detection). 
Figure 3 depicts an example of losses when training CAN to remove a category of subsidiary information from the code. 
During the first phase (left part of the figure), the encoder and the primary models are successfully trained. 
However, after freezing the encoder, loss of the subsidiary model does not decrease. 
In additional experiments where the encoder was not frozen when training the subsidiary information, the loss of the subsidiary task decreased successfully, but we did not achieve improvements in terms of EER. 

\begin{figure}[t]
\begin{center}
\centering
\includegraphics[width=\linewidth]{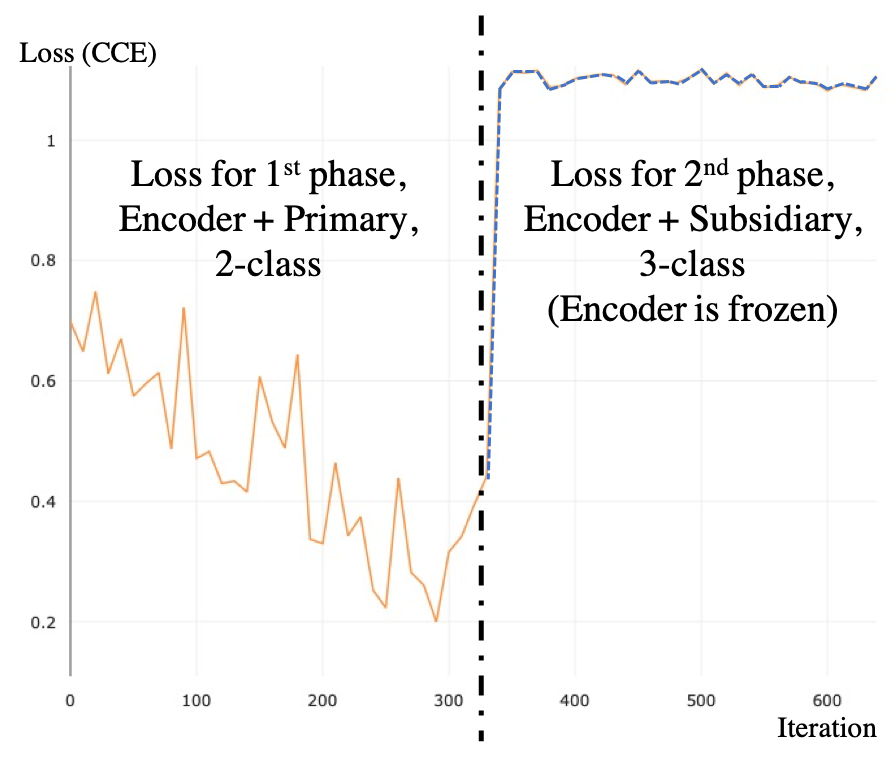} 
\caption{A typical example of the loss for removing a category of subsidiary information when adopting the CAN framework. Illustrated image is from the experiment removing `Replay Quality'. At the end of the first phase in CAN training, loss value is under 0.2. However, after freezing the encoder, loss for subsidiary task (classifying replay device quality here) does not decrease from approximately 1.2, similar to random classification.}
\label{fig:3}
\end{center}
\end{figure}

\subsection{Adding subsidiary information (MTL)}
Tables \ref{table:MTL} and \ref{table:modifiedMTL} describe the result of applying the MTL framework and the modified MTL framework, respectively. 
The performances on the development set (closed set configuration) improved in all five categories of subsidiary information, showing a relative error rate reduction (RER) greater than 30 \% (`Reverberation' in Table \ref{table:MTL}, `Attacker-to-Talker' in Table \ref{table:modifiedMTL}). 
However, in the evaluation set (open set configuration), we gained a minor performance improvement of RER 3 \% by using `Reverberation' and `Replay Quality' only. 

We analyze that the performance difference between the development set and the evaluation set is mainly due to two reasons. 
First, different configuration between the closed set the and open set have brought performance differences. 
Training and development set share common configurations for each subsidiary information. 
However, although five categories of subsidiary information is used, the evaluation set comprises open set replay configurations. 
Second, in each category of subsidiary information, the number of sub-category labels are too small and ambiguous. 
For example, for `Replay Quality' of the ASVspoof 2019 PA dataset, there exists only three different kinds of labels, `perfect', `high', and `low'.
We conclude that three sub-categories, with rather ambiguous labels are too small to generalize towards unknown open set configurations. 
With this conclusion, successful results using the ASVspoof 2017 evaluation dataset (open set) with the MTL framework are analyzed to have occurred owing to actual replay device labels. 

The modified MTL framework also demonstrated performance improvement only for the closed set configuration (development set). 
In the results of this study, using the MTL framework to add various subsidiary information did not show significant improvement in the evaluation set. 
However, significant improvement in the development set still demonstrates that when detailed subsidiary information labels are provided, adding these information can effectively improve the performance of the replay attack detection, and by using specifically labelled wide range of subsidiary labels, better generalization on open set configurations can be expected. 
Through experimental results, we concluded that various categories of subsidiary information can effectively aid in replay attack detection, but the process needs a wide range of specific labels for each category of subsidiary information.

\section{Conclusions}
\label{sec:5. future work}
In this study, we analyzed what sort of information is needed to conduct replay attack spoofing detection. 
For this purpose, we employed an E2E DNN and analyzed what information is included in the code embedding. 
Two frameworks, CAN and MTL were utilized to either subtract or add certain category of subsidiary information provided in the ASVspoof 2019 PA dataset to the code, respectively. 
Surprisingly, there was not enough relevant information to train the subsidiary model in the code extracted by the frozen encoder for all five categories of subsidiary information when the encoder has been trained using a binary classification scheme (bona-fide/replayed). 
Through addition of various categories of subsidiary information to the code, performance improvement was measured in closed set configuration, showing an RER over 30 \% for `Reverberation' and `Attacker-to-Talker'. 
We interpreted the above two results as following: the simple binary classification scheme was not appropriate for utilizing the subsidiary information which is helpful to replay spoofing attack detection.
However, in open set configuration, only minor performance improvements were observed. 
In our analysis, the minor performance improvements in open set are due to ambiguous, and small number of labels for each category of subsidiary information. 
As our future work, we intend to study different frameworks to include various categories of subsidiary information to the code to further benefit from additional information. 

\newpage
\bibliographystyle{IEEEbib}
\bibliography{refs}
\end{document}